\documentclass[twocolumn]{svjour3}
\usepackage[T2A]{fontenc}
\usepackage[russian,polish,english]{babel}
\usepackage{subcaption}
\usepackage{graphicx} 
\usepackage{url}
\usepackage{nameref}
\usepackage{hyperref}

\begin{document}

\title{Wikipedia Citations: Reproducible Citation Extraction from Multilingual Wikipedia}
\author{Natallia Kokash*\thanks{*Corresponding author} \and Giovanni Colavizza}

\institute{N. Kokash \at University of Amsterdam, Kloveniersburgwal 48, 1012 CX Amsterdam, The Netherlands
\email{natallia.kokash@gmail.com}
\and
G. Colavizza \at Department of Classical Philology and Italian Studies,
University of Bologna, Italy \email{giovanni.colavizza@unibo.it}
}

\authorrunning{N. Kokash, G. Colavizza}

\date{May 2024}

\maketitle

\begin{abstract}
Wikipedia is an essential component of the open science ecosystem, yet it is poorly integrated with academic open science initiatives. Wikipedia Citations is a project that focuses on extracting and releasing comprehensive datasets of citations from Wikipedia. A total of 29.3 million citations were extracted from English Wikipedia in May 2020. Following this one-off research project, we designed a reproducible pipeline that can process any given Wikipedia dump in the cloud-based settings. To demonstrate its usability, we extracted 40.6 million citations in February 2023 and 44.7 million citations in February 2024. Furthermore, we equipped the pipeline with an adapted Wikipedia citation template translation module to process multilingual Wikipedia articles in 15 European languages so that they are parsed and mapped into a generic structured citation template. This paper presents our open-source software pipeline to retrieve, classify, and disambiguate citations on demand from a given Wikipedia dump. 
    
\end{abstract}

\section{Introduction}
A \emph{citation} is a reference to a source such as a book, article, web page, or other published material that helps authors to support an argument, statement, fact, or acknowledge prior relevant discussions through their own intellectual work. Citations are an indispensable part of scientific publications as they help to trace the origins and evolution of ideas. Academic researchers take what is currently known and create new knowledge. In this process, citing sources is important for accumulating current state-of-the-art, emphasizing and getting credit for novel contributions, avoiding repetitive arguments and repetitive efforts, and helping anyone verify and validate the research results presented in any given scientific work.  

In the era of information explosion, we realized that scientific knowledge is hard to locate and process unless it is collected in online databases with additional search capabilities. Citation databases~\cite{citationIdx,citationDBs} are collections of referenced papers,  articles, books, and other materials entered into an online system in a structured and consistent way. All the information relating to a single document (author, title, publication details, abstract, and perhaps the full text) makes up the entry record for that document. Moreover, such records are connected to knowledge graphs that  
facilitate the search of relevant works, measuring the impact of a paper, an author or a journal, etc. General purpose citation databases such as Web of Science, Scopus, or Google Scholar cover a large number of recently published works. Field-specific databases such as PubMed and MathSciNet provide records on works from the most reputable journals in the respected fields.

Citations are increasingly used as performance indicators in research policy and within the research system. Yet, even with the presence of citation databases, the majority of scientific knowledge from specialized sources remains unavailable to the general public. The majority of people rely on search engines to find relevant information. Open online sources such as Wikipedia play a fundamental role in providing factual information on the Web. Wikipedia is one of the most visited websites in the world and is perceived as a reliable source of knowledge even for scientific research. 

In the scientific world researchers and authors generally follow well-established practices of citing relevant academic papers. Citations in other types of literature, and in particular, Wikipedia articles, may vary in quality and scope. The Wikipedia content quality, including the quality of its references, has been scrutinized in a number of previous studies~\cite{https://doi.org/10.1111/febs.15608,10.1162/qss_a_00171,https://doi.org/10.1002/asi.24723} and remains the subject of interest. However, any extensive study based on Wikipedia citations quickly becomes obsolete due to the constant evolution of its articles. Considering a large volume of periodically updated pages and attributes, we released a Wikipedia Citations extraction pipeline that allows anyone interested to obtain a comprehensive dataset of citations extracted from a chosen Wikipedia dump in a structured, unified and classified form. Our work is not limited to English Wikipedia, the pipeline includes a translation step that helps researchers and citation consolidation bodies to convert various citation templates from Wikicode into a common English-based data frame ready for automated processing and analysis.   

A crucial question to ask to improve Wikipedia’s verifiability standards, as well as to better understand its dominant role as a source of information, is the following: What sources are cited in Wikipedia? Our work is a direct continuation of the effort to extract and analyze a full set of English Wikipedia citations performed by Singh et al.~\cite{Singh2020}. The datasets extracted and classified in this work were used in a number of follow-up projects. Unlike previous Wikipedia-based citation studies, our goal was not to produce a one-time statistical analysis of a given snapshot but to develop a pipeline for any interested researcher to extract parsed citations mapped into a common template and enable Wikipedia citation integration with Open Science infrastructure. 

Answering the question of what exactly is cited in Wikipedia is challenging for a variety of reasons. First of all, editorial practices are not uniform, in particular across different language versions of Wikipedia. Secondly, while some citations contain stable identifiers (e.g., DOIs), others do not. As previous studies show~\cite{10.1371/journal.pone.0190046}, and we confirm in this paper, only a low fraction of all citation templates, e.g., around 5\% of citations in English Wikipedia, include DOIs. 

We expand upon previous work~\cite{Singh2020} by presenting a codebase for collecting citations from various language editions of Wikipedia. The extracted citations are translated to fit a common English schema and are categorized to identify reliable citation sources such as scientific journals, books, and news. We also present an improved lookup procedure to augment citations without identifiers if similar references are found in citation databases. 

This article is organized as follows. We start by describing our pipeline focusing on its three main steps: citation template harmonization, classification, and citation identifier lookup. We subsequently provide a description of the published data sets and compare our findings with similar Wikipedia citation evaluation methods.

\section{Workflow overview}

\begin{figure*}
    \centering
    \includegraphics[width=1\linewidth]{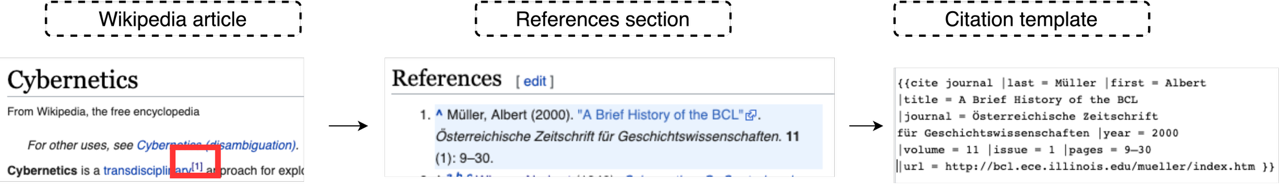}
    \caption{Wikipedia citation representation~\cite{Singh2020}.}
    \label{fig:wiki-citation}
\end{figure*}

Given a Wikipedia dump, our pipeline~\cite{WikiCite} retrieves citations and harmonizes citation templates, i.e., converts citations into a format that matches the same schema. Our work adopts the approach first developed by Singh et al.~\cite{Singh2020}. Figure~\ref{fig:wiki-citation} explains the technical representation of Wikipedia citations. 
A citation in Wikipedia is an abbreviated alphanumeric expression in the form of a Web link formatted as a subscript next to the relevant content of an article. The link directs a reader to a References section where a full textual citation identifying the source of the information is provided.
The citation text format depends on the essence of the cited resource (book, journal, web page, etc.) and is rendered automatically from a citation template chosen by the article author. 
A Wikipedia template is a page that is embedded into other pages to allow for the repetition of information. Citation templates are reusable pages specifically defined to embed citations.
They can be included in Wikicode (also known as Wikitext or Wiki markup), the markup language used to write Wikipedia pages, via double curly braces.    

\begin{figure*}
    \centering
    \includegraphics[width=1\linewidth]{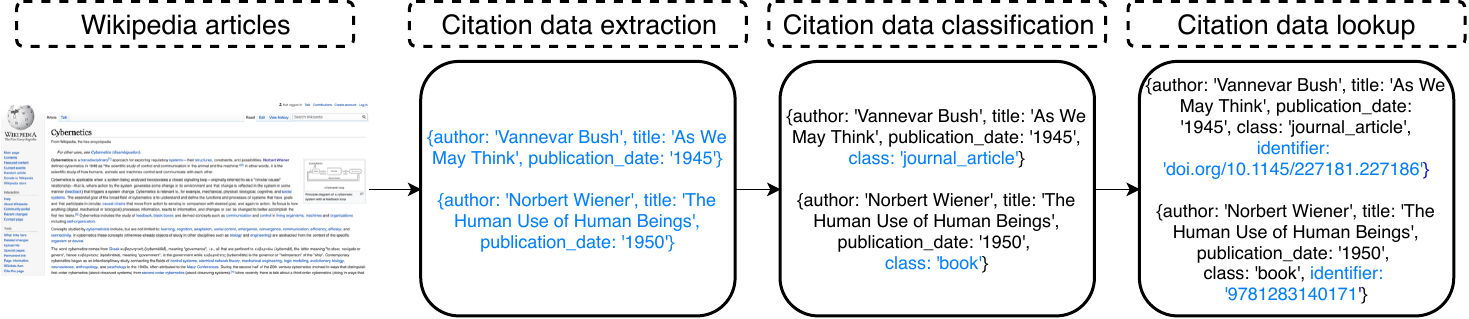}
    \caption{Citation extraction process~\cite{Singh2020}.}
    \label{fig:pipeline}
\end{figure*}

Figure~\ref{fig:pipeline} shows the basic steps of the citation extraction pipeline. 
XML Wikipedia dumps~\cite{WikimediaXML} provide raw Wikicode in which we locate all templates by extracting the content of double curly braces $\{\{...\}\}$ and $\langle ref \rangle$ tags. From these templates, we extract a template name and filter out templates unrelated to citation data. Wikipedia citation templates cover a much wider range of sources than those common for scientific literature, e.g., citing speeches, interviews, podcasts, press releases, court decisions, etc. For English Wikipedia, we process nearly 30 citation template types~\cite{Wiki-templates} which we parse with the extended version of the MediaWiki parser~\cite{mwparserfromhell}. Initially, the parser only supported 17 citation templates; support for an additional 18 of the most frequently used English templates was added~\cite{WikiCite-old}.

The resulting uniform key-value data set can easily be transformed into a tabular form for further processing. In particular, extracted in such a manner datasets with identifiers serve as input for our classification and lookup scripts, as well as are publicly released for anyone to analyze.   

The choice of a citation template by a Wikipedia article author cannot fully represent the nature of a cited source. Many templates such as \emph{citation}, \emph{cite web}, or their analogues in other languages are very generic and can be used to refer to scientific articles or arbitrary web pages with potentially non-factual content. Many Wikipedia content quality evaluation studies rely on the distinction between citations to scientific literature such as journal articles and conference proceedings or other content. To simplify such studies, we classify extracted citations equipped with known identifiers or URLs, labeling them as ``journal'', ``book'', ``news'', or ``other''. Furthermore, part of the citations which are likely to refer to books or journals (i.e., defined using corresponding Wikipedia templates) are augmented with identifiers located via Google Books and Crossref APIs. These steps, \emph{citation data classification} and \emph{citation data lookup}, are shown in Figure~\ref{fig:pipeline} and discussed in more detail in the next section.

Wikipedia is represented in 329 languages. Its authors can cite sources using either language-specific or English templates. Our main effort in releasing multi-lingual datasets was to assemble lists of citation templates for each language and convert relevant fields into a common English template. We started with known citation templates per language (typically covering books, journals, web pages, and news), and, in some cases, augmented these lists with additional frequently used templates (films, links, Web archives, etc.) which we were able to locate via the XML reference tags vs usage frequency dictionaries. 

\begin{figure*}
    \centering
    \includegraphics[width=1\linewidth]{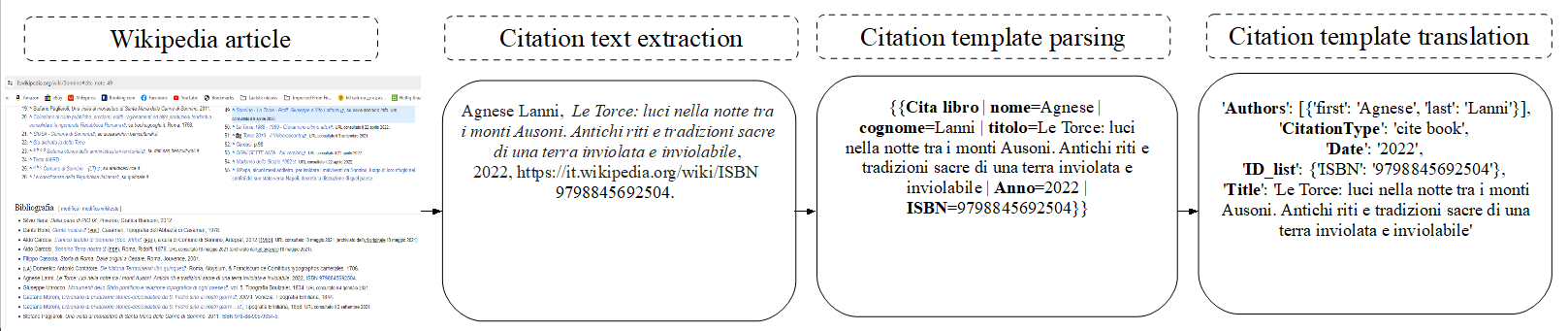}
    \caption{Citation translation process.}
    \label{fig:translation}
\end{figure*}

To extract citation data from multilingual dumps, we employed and extended a Wikipedia template translation module~\cite{Wiki-translator} which is part of the Wikipedia platform software. It is written in Lua and after resolving or substituting dependencies on other built-in modules, we integrated it into our Python-based pipeline using the Lupa library~\cite{lupa}. Similarly to English dumps (see Figure~\ref{fig:translation}), we extract all templates from Wikicode and select those that refer to language-specific or English citation templates. While English templates are parsed and mapped to uniform template key-value pairs directly, parsed language-specific templates are first translated into English. We only substitute template keys and leave their values unchanged. 

\section{Classification and lookup}

Along with information such as author(s), date of publication, title, and page numbers, citations may include unique identifiers depending on the type of work being referred to.
Citations of books often include an International Standard Book Number (ISBN).
Specific volumes, articles, or other identifiable parts of a periodical, may have an associated Serial Item and Contribution Identifier (SICI) or an International Standard Serial Number (ISSN). Electronic documents may have a Digital Object Identifier (DOI). Biomedical research articles may have a PubMed Identifier (PMID).

Similarly to~\cite{Singh2020}, we label citations with available identifiers:

\begin{itemize}
    \item Every citation with a PMC or PMID was labeled as a journal article.
    \item Every citation with a PMC, PMID, or DOI and using the citation template for journals and conferences was labeled as a journal article.
    \item Every citation that had an ISBN is labelled as a book.
    \item All citations with their URL top-level domain belonging to a given set of media news agencies are labeled as news.
\end{itemize}

Conceptually, our classification method is similar to the training set labeling for a neuron network classifier presented in~\cite{Singh2020}. While training a classifier is a valid approach for categorizing citation data, the preparation of the training dataset might be compromised by selection bias~\cite{selectionBias}. More specifically, the method selects citations with DOI, PMID, PMC, and ISBN identifiers and labels them as ``book'' or ``journal''; it also selects citations with top domains in URL that refer to a small list of news portals and labels them as ``web''. Then the classifier is trained and applied to label the whole dataset. However, this training dataset might not represent a randomized sampling from the whole dataset as variants of citations may not be included in the labeling process outlined above. Consequently, the high performance observed on the testing part of this same labeled dataset does not imply it will be performing equally well while labeling Wikipedia citations in general. Therefore, we release for public use~\cite{WikiCite-dataEnglish}\cite{WikiCite-dataMulti} only a conservatively labeled dataset with citations classified to ``book'', ``journal'', ``news'', and ``other'' categories using the aforementioned deterministic classifier. In the original work, the classifier was trained on a set of features: citation text, citation statement (the text preceding a citation in a Wikipedia article), Part of Speech (POS) tags, citation section, the order of the citation within the article, and the total number of words in the article. Our feature importance analysis showed that all features except for the citation text have a low impact on the classification accuracy. Nevertheless, we maintain the neural network-based classifier code as part of our codebase for maximum backward compatibility and future use.
 
To select citations that refer to the news, we collected news media links, 22646 unique website top domains, and social media page subdomains, from 4 sources:

\begin{enumerate}
    \item \url{mediabiasfactcheck.com}: The resource provides 9 categories of media sources, 5 of which are news portals with or without political bias (least, left, right, left center, and right center bias), and 4 categories with other sources (conspiracy-pseudo, questionable, satire, and science). We omit 4 latter categories as they are not suitable for classifying Wikipedia citations as ``news'' (while ``science'' domains are trustworthy, they list many journal websites - ideally, we would like to label Wikipedia citations with such URLs as ``journals''.
    \item \url{newspaperindex.com}: A list of the most important online newspapers and other news sources in all countries in the world. All the newspapers focus on general news, politics, debate, and the economy, and they are free to access online.
    \item \url{newsmedialists.com}: The resource lists links to newspapers, magazines, TV, and radio channels per country, as well as to their social network pages (Facebook, Instagram, Twitter, Pinterest, and YouTube).
    \item Projects \url{https://github.com/vegetable68/news_domain_labeled} and \texttt{us-news-domains} assembled a list of 3976 and 5000 news domains, respectively.
\end{enumerate}
We join top domains from the above resources into a set and classify a citation with a given URL as news if its top domain is in this set. 

A part of citations categorized as ``other'' due to the absence of identifiers may still represent references to books or journals. Citation data lookup is a process of searching for cited sources from Wikipedia in external citation databases. In our pipeline, we perform such a search in Crossref~\cite{crossref} and Google Books~\cite{googleBooks} databases for citation templates that imply a reference to a journal or a book (i.e, ``cite book'', ``cite journal'', ``cite encyclopedia'', ``cite proceedings'' in English Wikipedia) and augments the extracted dataset with retrieved identifiers if a match is found. The aforementioned citation databases return the best matches that may or may not coincide with the requested titles. Hence, we only consider the search successful if the returned title has a small editing distance~\cite{Levenshtein:65} from the requested title. 

The lookup with follow-up validation based on editing distance is a relatively slow process. To mitigate this step from potential failing (due to network issues, denial of service, wrong return format, etc.) on millions of requests, we process each parquet file separately and provide an option to save results in batches of a chosen size. 

\section{Wikipedia citation datasets}

\subsection{English Wikipedia}

The first release of the English Wikipedia Citations dataset from 2020~\cite{WikiCite-old} contains 4.0M citations to scholarly publications with known identifiers — including DOI, PMC, PMID, and ISBN — and further equip an extra 261K citations with DOIs from Crossref. Table~\ref{tab:eng-stats} shows updated data for English Wikipedia, which amounts for 5.0M citations to scientific journals and books in 2023 and 5.5M in 2024. These data were obtained using the same extraction pipeline and show the trend in Wikipedia growth, i.e., roughly a 10\% increase in the overall number of citations over a year.    

\begin{table*}
    \centering
    \caption{English Wikipedia citations}
    \begin{tabular}{ | l | r | r | r | r || r | r | r | r | r |}
    \hline
    Year & Citations  & Journals  & Books   & News  & Scientific & Reliable & Sci score & Sci score 2 & Rel score\\
    \hline
    2023 & 40.664.485 & 2.052.172 & 2.994.601 &  9.926.598  & 5.046.773 & 14.973.371 &  5.05 (4.55) & 12.41 & 36.82\\    
    2024 & 44.766.800 & 2.248.748 & 3.277.629 &  10.958.151 & 5.526.377 & 16.484.528 &  5.02        & 12.34 & 36.82\\
         & +10.0$\%$  &     9.6\% &    +9.4\% &    + 10.4\% &    +9.5\% &    +10.0\% & -0.03        & -0.07 &     0\\
    \hline
    \end{tabular}
    \label{tab:eng-stats}
\end{table*}

A \emph{sci score} is defined as the ratio of academic as opposed to non-academic references included in Wikipedia articles~\cite{Lewoniewski2023}. Along with the numbers of citations referring to published books, journals, and news, the table shows the sci score as a fraction of journal citations versus the whole number of extracted citations. In brackets next to the estimated sci score, 5\%, we provide the estimation based on the presence of DOI by Lewoniewski et al.~\cite{Lewoniewski2023}, also performed on 2023 English Wikipedia and equal to 4.55\%. As besides journals we also identify book citations, it may be interesting to see that the fraction of overall book and journal citations exceeds 12\% (sci score 2).    

With an effort to consolidate known lists of news media across the world, we were able to label a significant portion of Wikipedia citations as ``news''. Such citations generally can also be deemed trustworthy, assuming the majority of news media adhere to journalistic values and ensure that the information they distribute is accurate, fair, and thorough. Rel score in Table~\ref{tab:eng-stats} shows the fraction of all three classes of trustworthy information sources.   

While we only used 4 index types (DOI, PMC, PMID, and ISBN) to classify citations, primarily motivated by the question how 2023 and 2024 datasets differ from 2020, the following index types are also commonly used as annotations in Wikipedia citation templates: JSTOR, BIBCODE, ISMN, ISSN, SSRN, JFM, ARXIV, RFC, OL, LCCN, OCLC, ASIN, OSTI, MR, ZBL, USENETID, which potentially can be exploited for more thorough classification.  

Our experimentation with the lookup process showed that a significant number of Wikipedia citations are not recognized as scientific citations solely due to the absence of identifiers in their templates. In particular, 929.601 out of 2.445.913 English Wikipedia citations (38\%) we searched for in citation databases were augmented with unique identifiers via Google Books and Crossref APIs and hence refer to either books or journals. Together with these citations falling into journal and book classes, the Rel score 2 for English Wikipedia would amount to 14.7\%. 

Previously, we attempted to disambiguate references extracted from published books in Humanities~\cite{brill-pdfParser} using the same method and were able to retrieve identifiers roughly for 38\% of requested citations too. Although the estimation of citation database coverage was not the focus of any of these two projects, the fact that the disambiguation of two very different and substantial in size citation datasets yields similar outcomes in terms of the fraction of discovered entries, suggests that these two citation databases collectively cover not more than 40\% of cited sources that are likely to refer to scientific publications. Previous estimations of citation database coverage~\cite{martin-martin_google_2021} showed a higher percentage, although numbers significantly differed across services (Google Scholar, Microsoft Academic, Scopus, Dimensions, etc.) and coverage gaps in some areas, such as Physics and Humanities, were revealed. The likely explanation lies in the fact that this study evaluated coverage using a sample of citations taken from a seed sample of 2515 highly-cited documents listed in Google Scholar’s Classic Papers. While this is a viable approach for estimating the relative coverage of aforementioned databases, it may not be a fair dataset for answering the question of what overall percentage of cited scientific sources are present in existing citation databases. Our Wikipedia-based datasets provide interesting opportunities for such studies.  

\subsection{Multilingual Wikipedia}

\begin{table}
    \centering
    \caption{Multi-lingual Wikipedia citations}
    \begin{tabular}{ | l | r | r | r | r |}
        \hline
        Language & Size & Templates & Citations & \% \\
        \hline
        German     &   6.7GB & 36.513.001 & 4.854.945  & 13.3  \\
        French     &   5.9GB & 88.758.180 & 9.552.768  & 10.8  \\
        Russian    &   5.1GB & 37.486.749 & 7.437.100  & 19.84 \\
        Spanish    &   4.2GB & 29.436.355 & 6.918.442  & 23.5  \\
        Italian    &   3.6GB & 39.272.469 & 5.545.082  & 14.12 \\
        Polish     &   2.4GB & 24.347.212 & 4.744.158  & 19.5  \\
        Portuguese &   2.2GB & 22.579.438 & 4.775.025  & 21.15 \\
        Dutch      &   1.8GB & 18.812.097 &   566.549  &  3.0  \\
        Swedish    &   1.5GB & 21.900.024 & 3.802.416  & 17.36 \\
        Catalan    &   1.2GB & 11.905.510 & 2.239.714  & 18.81 \\
        Finnish    & 900.9MB &  7.433.203 & 1.697.731  & 22.84 \\ 
        Turkish    & 883.9MB &  8.890.506 & 1.993.177  & 22.42 \\
        Norwegian  & 763.7MB &  7.502.679 & 796.500    & 10.62 \\ 
        Danish     & 413.3MB &  3.176.509 & 437.239    & 13.76 \\
        \hline
    \end{tabular}
    \label{tab:multi-stats}
\end{table}

\begin{table*}
    \centering
    \caption{Multi-lingual Wikipedia reliable sources}
    \begin{tabular}{ | l | r | r | r | r || r | r | r | r | r|}
        \hline
        Language   &  Citations &  Books & Journals &     News & Scientific & Reliable & Sci score & Sci score 2 & Rel score \\
        \hline
        German     &  4.854.945 & 320.179 & 105.542 &   901.091 &   425.721 & 1.326.812 & 2.17 (1.88) &  8.77 & 27.33\\
        French     &  9.552.768 & 798.525 & 264.560 & 1.907.183 & 1.063.085 & 2.970.268 & 2.77 (1.84) & 11.13 & 31.09\\
        Russian    &  7.178.389 & 420.828 & 130.470 & 1.370.665 &   551.298 & 1.921.963 & 1.82 (1.84) &  7.68 & 25.84\\
        Spanish    &  6.918.442 & 522.910 & 213.767 & 1.699.396 &   736.677 & 2.436.073 & 3.09 (2.13) & 10.65 & 35.21\\
        Italian    &  5.545.082 & 384.816 & 128.366 &   917.517 &   513.182 & 1.430.699 & 2.31 (2.52) &  9.25 & 25.80\\
        Polish     &  4.744.158 & 463.783 &  95.988 &   513.006 &   559.771 & 1.072.777 & 2.11 (1.55) & 11.80 & 22.61\\
        Portuguese &  4.775.025 & 243.593 & 142.216 & 1.176.140 &   385.809 & 1.561.949 &  3.0 (2.61) &  8.08 & 32.71\\
        Dutch      &    566.549 &  27.074 &  12.706 &   114.110 &    39.780 &   133.890 & 2.24 (0.78) &  7.02 & 23.63\\
        Swedish    &  3.802.416 & 112.748 & 155.740 &   869.662 &   268.488 & 1.138.150 &  4.1 (1.44) &  7.06 & 29.93\\
        Catalan    &  2.239.714 & 261.779 & 105.125 &   423.241 &   366.904 &   790.145 &  4.7 (2.93) & 16.38 & 35.28\\
        Finnish    &  1.697.731 & 209.556 &  12.068 &   286.420 &   221.624 &   508.044 & 0.71 (0.59) & 13.05 & 29.92\\ 
        Turkish    &  1.993.177 &  85.079 &  56.202 &   339.122 &   141.281 &   480.403 & 2.82 (2.43) &  7.09 & 24.10\\
        Norwegian  &    796.500 &  43.314 &  12.373 &   151.780 &    55.687 &   207.467 & 1.55 (1.68) &  6.99 & 26.05\\ 
        Danish     &    437.239 &  23.303 &   7.522 &    70.760 &    30.825 &   101.585 & 1.72 (1.23) &  7.05 & 23.23\\
        \hline
    \end{tabular}
    \label{tab:multi-reliable}
\end{table*}

In the case of multilingual Wikipedia, we manually analyzed template frequency maps and selected which templates to parse for each language. Table~\ref{tab:multi-stats} shows the sizes of processed language dumps, the overall number of templates, the overall number of recognized citation templates, and the fraction of parsed citation templates versus the number of all template references, i.e., any inserts within $\langle ref \rangle$ and double curly braces $\{\{...\}\}.$  

Table~\ref{tab:multi-reliable} provides numbers related to citation classification in 14 languages other than English. Observe that our results correlate with the sci score reported in~\cite{Lewoniewski2023} (provided in brackets next to our score for convenience). Our estimation tends to be slightly higher for most languages. Apart from differences in the extraction process, this can be explained by a recent recognition of the importance of unique identifiers in Wikipedia citations. In contrast to our work, the authors do not map various multilingual citation templates to a common English data frame and estimate the number of scientific citations based on the simple presence of the DOI index in the citation text, which is the main focus of their work and is given for a full set of Wikipedia languages. 

Analogously to sci score, we compute the fraction of reliable citation sources, \emph{rel score}, which include references to published books, journals, and web resources at reputable news agency domains versus the whole number of citations. Catalan, Spanish, and French Wikipedia received higher scores on both such metrics.

Table~\ref{tab:multi-refs} presents counts for the five most common multi-lingual Wikipedia citation templates. For each language, native citation templates for journals, books, web pages, and news are among those used most. Often, their English counterparts are widely exploited, i.e., \emph{cite web} template appears among the most common citation templates in 9 out of 14 languages. 

\begin{table*}
    \caption{Five most common multi-lingual Wikipedia citation templates}
    \begin{subtable}{0.24\linewidth}   
        \centering
        \subcaption{German}
        \begin{tabular}{ | l | r |}
        \hline
        Template & Count \\
        \hline
         internetquelle & 2.842.270\\
         literatur & 907.044\\
         webarchiv & 740.540\\
          cite web & 284.434\\
         cite news &  31.502\\
        \hline
        \end{tabular}
    \end{subtable}
     \begin{subtable}{0.24\linewidth}
        \centering
        \subcaption{French}
        \begin{tabular}{ | l | r |}
        \hline
        Template & Count \\
        \hline
           lien web & 5.976.136 \\
            ouvrage & 1.502.055 \\
            article & 1.266.557 \\
           citation & 806.564 \\
        cite report &    386 \\
        \hline
        \end{tabular}
    \end{subtable}
    \begin{subtable}{0.24\linewidth}
        \centering
        \subcaption{Russian}
        \begin{tabular}{ | l | r |}
        \hline
        Template & Count \\
        \hline
            cite web & 3.175.967\\
            wayback  & 690.092\\
            \foreignlanguage{russian}{книга} & 546.076\\
            \foreignlanguage{russian}{статья}  & 278.650\\
          cite news  & 189.256\\
        \hline
        \end{tabular}
    \end{subtable}    
    \begin{subtable}{0.24\linewidth}
        \centering
        \subcaption{Spanish}
        \begin{tabular}{ | l | r |}
        \hline
        Template & Count \\
        \hline
           cita web & 4.076.779 \\
         cita libro & 711.873 \\
       cita noticia & 514.859 \\
           cite web & 504.741 \\
   cita publicación & 494.706 \\
        \hline
        \end{tabular}
    \end{subtable}  
    \vspace{8px}
    \begin{subtable}{0.27\linewidth}
        \centering
        \subcaption{Italian}
        \begin{tabular}{ | l | r |}
            \hline
            Template & Count \\
            \hline
                   cita web & 3.735.469\\
                 cita libro &  840.491\\
                  cita news & 476.739\\
         cita pubblicazione & 318.901\\
            cita conferenza &   1.648\\
            \hline
        \end{tabular}
    \end{subtable}
 \begin{subtable}{0.23\linewidth}
        \centering
        \subcaption{Polish}
        \begin{tabular}{ | l | r |}
        \hline
        Template & Count \\
        \hline
            cytuj stron\c e & 2530191 \\
                   cytuj & 1351399 \\
           cytuj ksi\c a\.zk\c e & 624640 \\
             cytuj pismo & 237926 \\
                cite web &      2 \\
        \hline
        \end{tabular}
    \end{subtable}    
    \begin{subtable}{0.25\linewidth}
        \centering
        \subcaption{Portuguese}
        \begin{tabular}{ | l | r |}
        \hline
        Template & Count \\
        \hline
          citar web  & 3.446.748 \\
        citar livro  &   383.034 \\
    citar periódico  &   291.663 \\
               link  &   211.959 \\
        citar jornal &   199.418 \\
        \hline
        \end{tabular}
    \end{subtable}  
    \begin{subtable}{0.23\linewidth}
        \centering
        \subcaption{Dutch}
        \begin{tabular}{ | l | r |}
        \hline
        Template & Count \\
        \hline
            citeer web& 431.570\\
              cite web& 57.149\\
           citeer boek& 41.465\\
             cite news& 11.489\\
        citeer journal& 10.718\\
        \hline
        \end{tabular}
    \end{subtable}
    \vspace{8px}
    \begin{subtable}{0.24\linewidth}
        \centering
        \subcaption{Swedish}
        \begin{tabular}{ | l | r |}
        \hline
        Template & Count \\
        \hline
             webbref & 3.176.499\\
             bokref & 228.791 \\
      tidskriftsref & 169.625 \\
            wayback & 166.952 \\
           cite web &  31.073 \\
        \hline
        \end{tabular}
    \end{subtable}   
    \begin{subtable}{0.24\linewidth}
        \centering
        \subcaption{Finnish}
        \begin{tabular}{ | l | r |}
        \hline
        Template & Count \\
        \hline
        verkkoviite & 1.175.985 \\
         kirjaviite & 267.953 \\
            wayback & 152.801 \\
         lehtiviite &  66.079 \\
           cite web &  24.799 \\
        \hline
        \end{tabular}
    \end{subtable}   
    \begin{subtable}{0.24\linewidth}
        \centering
        \subcaption{Turkish}
        \begin{tabular}{ | l | r |}
        \hline
        Template & Count \\
        \hline
             web kaynağı & 1.264.127 \\
                webarşiv & 356.897 \\
           haber kaynağı & 149.095 \\
           kitap kaynağı & 126.230 \\
           dergi kaynağı &  94.657 \\
        \hline
        \end{tabular}
    \end{subtable}
     \begin{subtable}{0.24\linewidth}
        \centering
        \subcaption{Norwegian}
        \begin{tabular}{ | l | r |}
        \hline
        Template & Count \\
        \hline
         kilde www & 374.391 \\
           wayback & 124.632 \\
          cite web & 116.100 \\
         kilde bok & 56.554 \\
        kilde avis & 43.818 \\
        \hline
        \end{tabular}
    \end{subtable}    
    \vspace{8px}
    \begin{subtable}{0.25\linewidth}
        \centering
        \subcaption{Catalan}
        \begin{tabular}{ | l | r |}
        \hline
        Template & Count \\
        \hline
           ref-web & 1.185.404\\
        ref-llibre & 389.447\\
    ref-publicació & 265.016\\
        webarchive & 211.238\\
       ref-notícia & 118.187\\
        \hline
        \end{tabular}
    \end{subtable}  
    \begin{subtable}{0.25\linewidth}
        \centering
        \subcaption{Danish}
        \begin{tabular}{ | l | r |}
        \hline
        Template & Count \\
        \hline
           cite web & 193.539 \\
         webarchive & 109.005 \\
          cite news & 42.853 \\
          cite book & 25.639 \\
              kilde & 20.804 \\
        \hline
        \end{tabular}
    \end{subtable}    
    \label{tab:multi-refs}
\end{table*}

Our released datasets~\cite{WikiCite-dataEnglish,WikiCite-dataMulti} include the following columns:
\begin{itemize}
    \item \texttt{type of citation}: Wikipedia template type used to define the citation, e.g., \emph{cite book}.
    \item \texttt{page title}: title of the Wikipedia article from which the citation was extracted.
    \item \texttt{title}: source title, e.g., title of the book, newspaper article, etc.
    \item \texttt{url}: link to the source, e.g., the web page where the news article was published, description of the book at the publisher's website, online library web page, etc.
    \item \texttt{tld}: top link domain extracted from the URL, e.g., \texttt{bbc} for \url{https://www.bbc.co.uk/}. 
    \item \texttt{authors}: list of article or book authors, if available.
    \item \texttt{ID list}: list of publication identifiers mentioned in the citation, e.g., DOI, ISBN, etc.
    \item \texttt{citation}: citation text as used in Wikipedia code 
    \item \texttt{actual label}: ``book'', ``journal'', ``news'', or ``other'' label assigned based on the analysis of citation identifiers or top link domain.    
    \item \texttt{acquired ID list}: identifiers located by the lookup process for selected citation types.
\end{itemize}

A much wider set of citation template properties is retrieved, parsed, and mapped into a uniform format by the pipeline, e.g., publication date, publication place, publisher name, issue, volume, pages, etc. These extended datasets can be shared on demand for the 15 February 2024 dumps discussed in this paper. The code is open-source and we provide detailed instructions on how to run extraction, classification, and lookup scripts on Google Cloud Dataproc, a fast, easy-to-use, and fully managed service for running Apache Spark, see documentation in the project repository~\cite{WikiCite,WikiCite-multi}.

\section{Related work}

Open science~\cite{openScience} is the movement to make scientific research and its dissemination accessible to all levels of society. Wikipedia is a popular open science platform that over years has become an integral part of scholarly communication. Kousha and Thelwall~\cite{https://doi.org/10.1002/asi.23694} assessed the value of Wikipedia citations for academic articles and books, discovering that 5\% of articles and 33\% of monographs have at
least one citation from Wikipedia. On the other hand, Wikipedia helps popularize academic research. Various studies have shown that a high portion of citations to sources in Wikipedia refer to scientific or scholarly literature~\cite{Nielsen-17}. The literature cited in Wikipedia has been found to correlate positively with a journal’s popularity, its impact factor, and open access policy~\cite{10.1162/qss_a_00226,https://doi.org/10.1002/asi.23694,https://doi.org/10.1002/asi.23833}. A clear influence of Wikipedia on scientific research has been found~\cite{Thompson2018ScienceIS}, despite a general lack of reciprocity in acknowledging it as a source of information from the scientific literature~\cite{https://doi.org/10.1002/asi.23691,Tomaszewski-MacDonald-16}. 
Many efforts are dedicated to the task of expanding and improving Wikipedia’s verifiability through citations to reliable sources~\cite{Fetahu-17,Fetahu-19}. Citations in Wikipedia are also useful for users browsing low-quality or underdeveloped articles, as they allow them to look for information outside of the platform~\cite{PiccardiEtAl2020}. 

Wikipedia is instrumental in providing access to scientific information and in fostering the public understanding of science~\cite{Heilman-11,wiki-humanities}. According to 2017 study~\cite{Teplitskiy2015AmplifyingTI}, open-access journals have a 47\% higher chance of being cited in Wikipedia. Our datasets of classified citations may assist researchers in analyzing the impact and distribution of scientific knowledge via Wikipedia, e.g., provided a list of open-access journals, one can measure their portion among citations in various Wikipedia editions. Due to the importance of citations for many studies around Wikipedia as a crucial part of open science infrastructure, there is an ongoing effort to produce full sets of Wikipedia citations. Zagovora et al.~\cite{10.1162/qss_a_00171} present a data set that contains individual revision histories of all Wikipedia references ever created in the English Wikipedia until June 2019. Regardless of how complete the extracted set is, with millions of Wikipedia edits per year, no set can be considered final. Therefore we wanted to equip researchers interested in Wikipedia studies with an open and easy-to-use tool to collect citations.  

Early studies of what scientific and scholarly literature is cited in Wikipedia point to relatively low coverage, indicating that between 1$\%$ and 5$\%$ of all published journal articles are cited in Wikipedia~\cite{Pooladian2017MethodologicalII,10.1162/qss_a_00171}. The authors of this study found a persistent increase of references equipped with some form of document identifier over time, they underline how relying on references with document identifiers is still not sufficient to capture all relevant publications cited from Wikipedia. Our estimation of scientific citations is higher and consistent among multi-lingual dumps. We show that via the augmentation of citations this percentage can be significantly higher as identifiers are often omitted in citations. Moreover, even the look-up process does not solve the problem of recognizing scientific references completely. While citation database coverage estimations differ across studies and domains~\cite{martin-martin_google_2021,10.1162/qss_a_00112}, based on our experience with English Wikipedia citation augmentation discussed in this paper, less than half of likely scientific citations (more precisely, 38\%) are found via Google Books and Crossref search (which is remarkably close to the fraction of disambiguated citations from the Brill's dataset~\cite{brill-kg} in our previous project).  

The majority of Wikipedia citation studies are done on its English version only. Lewoniewski et at al. analyses the use of common references in several Wikipedia language editions: English, German, French, Russian, Polish, Ukrainian, and Belarussian~\cite{Lewoniewski17}. Later the same authors present descriptive analysis of DOI-containing references in 310 Wikipedia languages~\cite{Lewoniewski2023}. However, these studies do not translate and align citations across languages. This part of our contribution opens opportunities for more in-depth studies of multilingual Wikipedia citation structures and trends.

Maggio et al.~\cite{10.1371/journal.pone.0190046} establish benchmarks for the relative distribution and click rate of citations with DOI from English Wikipedia, with a focus on medical citations. Arroyo-Machado et al.~\cite{10.1162/qss_a_00226} construct an open knowledge graph for the large-scale analysis of the English Wikipedia metrics. A complete relational dataset of English Wikipedia is built. Pooladian and Borrego~\cite{Pooladian2017} evaluate the impact of research papers by their exposure to a wider audience beyond the academic community. The paper evaluates the Wikipedia references to papers in Information Science published between 2001 and 2010 and concludes that less than 3\% of articles in the sample were cited and that nearly one-third of the Wikipedia citations link to open-access sources. Jemielniak et al.~\cite{Jemielniak19} determine the ranking of the most cited journals by their representation in the English-language medical pages of Wikipedia.
Torres-Salinas et al.~\cite{wiki-humanities} study how scientific knowledge is presented in Wikipedia articles in the field of Humanities, finding out that the citation average for Humanities articles in Wikipedia is lower than the general.
Colavizza~\cite{10.1162/qss_a_00080} evaluates Wikipedia's rapid response to Covid-19 research observing that almost 2\% from  160.000 scientific articles were already cited in Wikipedia in a few months after the start of the pandemic. This supports our view that any snapshot-based evaluation gets obsolete quickly and continuous integration of citation sources with academic infrastructure is needed.

Nicholson et al.~\cite{https://doi.org/10.1111/febs.15608} analyzed scientific articles referenced in the English Wikipedia to see how they had been cited in the scientific literature. The authors find that around 3\% of citations are used to provide supporting evidence, 0.35\% to provide contradicting evidence, and 96.7\% mention a study without indicating that they provide supporting or contradicting evidence. Zheng et al.~\cite{https://doi.org/10.1002/asi.24723} assess gender- and country-based biases in Wikipedia citation practices. The research has found that publications by women are cited less than those by men. Scholarly publications by authors affiliated with non-Anglosphere countries are also disadvantaged in getting cited by Wikipedia. The level of gender- or country-based inequalities varies by research field, and the gender-country intersectional bias is prominent in math-intensive STEM fields. 

Among MediaWiki parser implementations that work on XML dumps~\cite{WikimediaXML} only a few offer complete or near complete coverage and are regularly maintained~\cite{wiki-parsers}. Recently a Python library for parsing and mining metadata from the Wikipedia HTML dumps~\cite{WikimediaHTML} has been made available~\cite{mwparserfromhtml}. Besides using the HTML dumps, users can also use the Wikipedia API to obtain the HTML of a particular article from their title and parse the HTML string with this library.  

\section{Conclusion and Future Work}
This contribution is a direct continuation of the previous effort~\cite{Singh2020,Lewoniewski2023} by the community to study, use, maintain, and expand work on citations from Wikipedia. 

We published the Wikipedia Citations datasets in 15 languages. Citations in multilingual datasets are translated and harmonized to match a common template. We used persistent identifiers such as DOIs and ISBNs and top domains extracted from URL references whenever available to classify citations into four categories: ``books'', ``journals'', ``news'', and ``other''. To identify news references, we assembled a comprehensive list of news media agencies worldwide. Furthermore, we refined a mechanism to annotate citations without persistent identifiers such from Google Books and Crossref citation databases. 

We collected basic statistics about 2024 datasets and compared them with available data from previous years. Our results show that e.g., the number of Wikipedia citations increased by 10\% since 2023, and the sci score of the majority of the processed language editions, i.e., the percentage of scientific journal citations, has improved reaching 2.61\% vs 1.88\% previously on average for the 15 languages we analyzed. We also release all our code to extend upon our work and update the datasets in the future, equipped with detailed instructions on how to run it reliably on Google Cloud infrastructure. We believe that this code is easy to extend for processing of other language editions by specifying language-specific citation template names and mapping their properties to English counterparts (``authors'', ``title'', etc.), fostering cross-cultural research on Wikipedia content. 

\section*{Acknowledges}
{\small 
This research was supported in part by the University of Amsterdam Data Science Centre.
}

\bibliographystyle{spmpsci}
\bibliography{main}

\begin{thebibliography}{10}
\providecommand{\url}[1]{{#1}}
\providecommand{\urlprefix}{URL }
\expandafter\ifx\csname urlstyle\endcsname\relax
  \providecommand{\doi}[1]{DOI~\discretionary{}{}{}#1}\else
  \providecommand{\doi}{DOI~\discretionary{}{}{}\begingroup
  \urlstyle{rm}\Url}\fi

\bibitem{wiki-parsers}
Alternative parsers (2024).
\newblock \urlprefix\url{https://www.mediawiki.org/wiki/Alternative_parsers}

\bibitem{10.1162/qss_a_00226}
Arroyo-Machado, W., Torres-Salinas, D., Costas, R.: {Wikinformetrics:
  Construction and description of an open Wikipedia knowledge graph data set
  for informetric purposes}.
\newblock Quantitative Science Studies \textbf{3}(4), 931--952 (2022).
\newblock \doi{10.1162/qss_a_00226}.
\newblock \urlprefix\url{https://doi.org/10.1162/qss\_a\_00226}

\bibitem{lupa}
Behnel, S.: {Lupa: Python wrapper around Lua and LuaJIT} (2024).
\newblock \urlprefix\url{https://pypi.org/project/lupa/}

\bibitem{citationIdx}
Castanha, Hj\o{}rland, B., Gutierres, R.C., de~Ara\'{u}jo, P.C.: Citation
  indexing and indexes.
\newblock Knowledge Organization \textbf{48}(1), 72--101 (2021).
\newblock \doi{10.5771/0943-7444-2021-1-72}

\bibitem{10.1162/qss_a_00080}
Colavizza, G.: {COVID-19} research in {Wikipedia}.
\newblock Quantitative Science Studies \textbf{1}(4), 1349--1380 (2020).
\newblock \doi{10.1162/qss_a_00080}.
\newblock \urlprefix\url{https://doi.org/10.1162/qss\_a\_00080}

\bibitem{crossref}
{Crossref REST API Documentation} (2016).
\newblock
  \urlprefix\url{https://www.crossref.org/documentation/retrieve-metadata/rest-api/}

\bibitem{WikimediaXML}
Dumps, W.: {Wikimedia Downloads} (2024).
\newblock \urlprefix\url{https://dumps.wikimedia.org/backup-index.html}

\bibitem{WikimediaHTML}
Dumps, W.: {Wikimedia Enterprise HTML Dumps} (2024).
\newblock \urlprefix\url{https://dumps.wikimedia.org/other/enterprise_html/}

\bibitem{Fetahu-17}
Fetahu, B., Markert, K., Anand, A.: {Fine Grained Citation Span for References
  in Wikipedia}.
\newblock In: Proceedings of the 2017 Conference on Empirical Methods in
  Natural Language Processing, pp. 1990--1999. Association for Computational
  Linguistics (2017).
\newblock \doi{10.18653/v1/D17-1212}

\bibitem{googleBooks}
{Google Books APIs: Volume} (2012).
\newblock
  \urlprefix\url{https://developers.google.com/books/docs/v1/reference/volumes}

\bibitem{openScience}
Grand, A.: Open science.
\newblock Journal of Science Communication \textbf{14} (2015).
\newblock \doi{10.22323/2.14040302}

\bibitem{Heilman-11}
Heilman, J., Kemmann, E., Bonert, M., Chatterjee, A., Ragar, B., Beards, G.,
  Iberri, D., Harvey, M., Thomas, B., Stomp, W., Martone, M., Lodge, D.,
  Vondracek, A., De~Wolff, J., Liber, C., Grover, S., Vickers, T., Mesko, B.,
  Laurent, M.: Wikipedia: A key tool for global public health promotion.
\newblock Journal of medical Internet research \textbf{13}, e14 (2011).
\newblock \doi{10.2196/jmir.1589}

\bibitem{https://doi.org/10.1002/asi.23691}
Jemielniak, D., Aibar, E.: Bridging the gap between {Wikipedia} and academia.
\newblock Journal of the Association for Information Science and Technology
  \textbf{67}(7), 1773--1776 (2016).
\newblock \doi{https://doi.org/10.1002/asi.23691}.
\newblock
  \urlprefix\url{https://asistdl.onlinelibrary.wiley.com/doi/abs/10.1002/asi.23691}

\bibitem{Jemielniak19}
Jemielniak, D., Masukume, G., Wilamowski, M.: The most influential medical
  journals according to {Wikipedia}: Quantitative analysis.
\newblock Journal of Medical Internet Research \textbf{21}, e11429 (2019).
\newblock \doi{10.2196/11429}

\bibitem{WikiCite}
Kokash, N.: {Extraction and classification of references from Wikipedia
  articles} (2024).
\newblock \urlprefix\url{https://github.com/albatros13/wikicite}

\bibitem{WikiCite-multi}
Kokash, N.: {Extraction and classification of references from Wikipedia
  articles in multiple languages} (2024).
\newblock \urlprefix\url{https://github.com/albatros13/wikicite/tree/multilang}

\bibitem{WikiCite-dataEnglish}
Kokash, N., Colavizza, G.: {A Comprehensive Dataset of Classified Citations
  with Identifiers from English Wikipedia (2024)} (2024).
\newblock \urlprefix\url{https://zenodo.org/records/10782978}

\bibitem{WikiCite-dataMulti}
Kokash, N., Colavizza, G.: {A Comprehensive Dataset of Classified Citations
  with Identifiers from Multilingual Wikipedia (2024)} (2024).
\newblock \urlprefix\url{https://zenodo.org/records/11210434}

\bibitem{brill-pdfParser}
Kokash, N., Romanello, M., Suyver, E., Colavizza, G.: From books to knowledge
  graphs.
\newblock Journal of Data Mining and Digital Humanities  (2023)

\bibitem{brill-kg}
Kokash, N., Romanello, M., Suyver, E., Colavizza, G.: The {Brill} knowledge
  graph: A database of bibliographic references and index terms extracted from
  books in {Humanities and Social Sciences}.
\newblock Research Data Journal for the Humanities and Social Sciences pp. 1 --
  21 (2024).
\newblock \doi{10.1163/24523666-bja10036}.
\newblock
  \urlprefix\url{https://brill.com/view/journals/rdj/aop/article-10.1163-24523666-bja10036/article-10.1163-24523666-bja10036.xml}

\bibitem{https://doi.org/10.1002/asi.23694}
Kousha, K., Thelwall, M.: {Are Wikipedia citations important evidence of the
  impact of scholarly articles and books?}
\newblock Journal of the Association for Information Science and Technology
  \textbf{68}(3), 762--779 (2017).
\newblock \doi{https://doi.org/10.1002/asi.23694}.
\newblock
  \urlprefix\url{https://asistdl.onlinelibrary.wiley.com/doi/abs/10.1002/asi.23694}

\bibitem{mwparserfromhell}
Kurtovic, B.: {A Python parser for MediaWiki wikicode (mwparserfromhell)}
  (2024).
\newblock \urlprefix\url{https://github.com/earwig/mwparserfromhell}

\bibitem{Levenshtein:65}
Levenshtein, V.I.: {Binary Codes Capable of Correcting Spurious Insertions and
  Deletions of Ones}.
\newblock Probl. Inf. Transm. \textbf{1}(1), 8–--17 (1965)

\bibitem{Lewoniewski2023}
Lewoniewski, W., W\c{e}cel, K., Abramowicz, W.: Understanding the use of
  scientific references in multilingual {Wikipedia} across various topics.
\newblock Procedia Computer Science \textbf{225}, 3977--3986 (2023).
\newblock \doi{https://doi.org/10.1016/j.procs.2023.10.393}.
\newblock
  \urlprefix\url{https://www.sciencedirect.com/science/article/pii/S187705092301551X}.
\newblock 27th Int. Conference on Knowledge Based and Intelligent Information
  and Engineering Sytems (KES 2023)

\bibitem{Lewoniewski17}
Lewoniewski, W., W{\c{e}}cel, K., Abramowicz, W.: Analysis of References Across
  {Wikipedia} Languages, 756 edn., pp. 561--573.
\newblock Springer (2017).
\newblock \doi{10.1007/978-3-319-67642-5_47}

\bibitem{10.1371/journal.pone.0190046}
Maggio, L.A., Willinsky, J.M., Steinberg, R.M., Mietchen, D., Wass, J.L., Dong,
  T.: Wikipedia as a gateway to biomedical research: The relative distribution
  and use of citations in the {English Wikipedia}.
\newblock PLOS ONE \textbf{12}(12), 1--12 (2017).
\newblock \doi{10.1371/journal.pone.0190046}.
\newblock \urlprefix\url{https://doi.org/10.1371/journal.pone.0190046}

\bibitem{martin-martin_google_2021}
Mart\'{i}n-Mart\'{i}n, A., Thelwall, M., Orduna-Malea, E., Delgado
  L\'{o}pez-C\'{o}zar, E.: {Google Scholar, Microsoft Academic, Scopus,
  Dimensions, Web of Science, and OpenCitations’ COCI}: a multidisciplinary
  comparison of coverage via citations.
\newblock Scientometrics \textbf{126}(1), 871--906 (2021).
\newblock \doi{10.1007/s11192-020-03690-4}.
\newblock \urlprefix\url{https://link.springer.com/10.1007/s11192-020-03690-4}

\bibitem{citationDBs}
Neuhaus, C., Daniel, H.D.: Data sources for performing citation analysis: An
  overview.
\newblock Journal of Documentation \textbf{64}, 193--210 (2008).
\newblock \doi{10.1108/00220410810858010}

\bibitem{https://doi.org/10.1111/febs.15608}
Nicholson, J.M., Uppala, A., Sieber, M., Grabitz, P., Mordaunt, M., Rife, S.C.:
  Measuring the quality of scientific references in {Wikipedia}: an analysis of
  more than 115m citations to over 800 000 scientific articles.
\newblock The FEBS Journal \textbf{288}(14), 4242--4248 (2021).
\newblock \doi{https://doi.org/10.1111/febs.15608}.
\newblock
  \urlprefix\url{https://febs.onlinelibrary.wiley.com/doi/abs/10.1111/febs.15608}

\bibitem{Nielsen-17}
Nielsen, F., Mietchen, D., Willighagen, E.: Scholia, scientometrics and
  wikidata.
\newblock In: The Semantic Web: ESWC 2017 Satellite Events, LNCS, pp. 237--259
  (2017).
\newblock \doi{10.1007/978-3-319-70407-4_36}

\bibitem{PiccardiEtAl2020}
Piccardi, T., Redi, M., Colavizza, G., West, R.: Quantifying engagement with
  citations on {Wikipedia}.
\newblock In: WWW '20: Proceedings of The Web Conference 2020, pp. 2365--2376
  (2020).
\newblock \doi{10.1145/3366423.3380300}

\bibitem{Pooladian2017MethodologicalII}
Pooladian, A., Borrego, {\'A}.: Methodological issues in measuring citations in
  {Wikipedia}: a case study in library and information science.
\newblock Scientometrics \textbf{113}, 455 -- 464 (2017).
\newblock \urlprefix\url{https://api.semanticscholar.org/CorpusID:255007038}

\bibitem{Pooladian2017}
Pooladian, A., Borrego, {\'A}.: Methodological issues in measuring citations in
  {Wikipedia}: a case study in {Library and Information Science}.
\newblock Scientometrics \textbf{113}(1), 455--464 (2017).
\newblock \doi{10.1007/s11192-017-2474-z}.
\newblock \urlprefix\url{https://doi.org/10.1007/s11192-017-2474-z}

\bibitem{Fetahu-19}
Redi, M., Fetahu, B., Morgan, J., Taraborelli, D.: Citation needed: A taxonomy
  and algorithmic assessment of {Wikipedia's} verifiability.
\newblock In: WWW '19: The World Wide Web Conference, pp. 1567--1578 (2019).
\newblock \doi{10.1145/3308558.3313618}

\bibitem{Singh2020}
Singh, H., West, R., Colavizza, G.: {Wikipedia citations: A comprehensive data
  set of citations with identifiers extracted from English Wikipedia}.
\newblock Quantitative Science Studies \textbf{2}, 1--19 (2020).
\newblock \doi{10.1162/qss_a_00105}

\bibitem{WikiCite-old}
Singh, H., West, R., Colavizza, G.: {Wikipedia Citations: A comprehensive
  dataset of citations with identifiers extracted from English Wikipedia}
  (2020)

\bibitem{https://doi.org/10.1002/asi.23833}
Sugimoto, C.R., Work, S., Larivi{\'e}re, V., Haustein, S.: Scholarly use of
  social media and altmetrics: A review of the literature.
\newblock Journal of the Association for Information Science and Technology
  \textbf{68}(9), 2037--2062 (2017).
\newblock \doi{https://doi.org/10.1002/asi.23833}.
\newblock
  \urlprefix\url{https://asistdl.onlinelibrary.wiley.com/doi/abs/10.1002/asi.23833}

\bibitem{Teplitskiy2015AmplifyingTI}
Teplitskiy, M., Lu, G., Duede, E.: Amplifying the impact of open access:
  Wikipedia and the diffusion of science.
\newblock Journal of the Association for Information Science and Technology
  \textbf{68} (2015).
\newblock \urlprefix\url{https://api.semanticscholar.org/CorpusID:10220883}

\bibitem{Wiki-translator}
The Wikimedia~Foundation, I.: {Module:CS1 translator} (2024).
\newblock \urlprefix\url{https://en.wikFipedia.org/wiki/Module:CS1_translator}

\bibitem{Wiki-templates}
The Wikimedia~Foundation, I.: {Wikipedia: Citation templates} (2024).
\newblock
  \urlprefix\url{https://en.wikipedia.org/wiki/Wikipedia:Citation_templates}

\bibitem{Thompson2018ScienceIS}
Thompson, N.C., Hanley, D.: Science is shaped by {Wikipedia}: Evidence from a
  randomized control trial.
\newblock IRPN: Institutional  (2018).
\newblock \urlprefix\url{https://api.semanticscholar.org/CorpusID:30918097}

\bibitem{Tomaszewski-MacDonald-16}
Tomaszewski, R., MacDonald, K.: A study of citations to {Wikipedia} in
  scholarly publications.
\newblock Science \& Technology Libraries \textbf{35}, 1--16 (2016).
\newblock \doi{10.1080/0194262X.2016.1206052}

\bibitem{wiki-humanities}
Torres-Salinas, D., Romero-Frías, E., Arroyo-Machado, W.: {Mapping the
  backbone of the Humanities through the eyes of Wikipedia}.
\newblock Journal of Informetrics \textbf{13}(3), 793--803 (2019).
\newblock \doi{10.1016/j.joi.2019.07.002}.
\newblock
  \urlprefix\url{https://ideas.repec.org/a/eee/infome/v13y2019i3p793-803.html}

\bibitem{10.1162/qss_a_00112}
Visser, M., van Eck, N.J., Waltman, L.: {Large-scale comparison of
  bibliographic data sources: Scopus, Web of Science, Dimensions, Crossref, and
  Microsoft Academic}.
\newblock Quantitative Science Studies \textbf{2}(1), 20--41 (2021).
\newblock \doi{10.1162/qss_a_00112}.
\newblock \urlprefix\url{https://doi.org/10.1162/qss\_a\_00112}

\bibitem{mwparserfromhtml}
{Wikimedia Enterprise}: {A Python library for parsing and mining metadata from
  the Enterprise HTML Dumps (mwparserfromhtml)} (2024).
\newblock \urlprefix\url{https://pypi.org/project/mwparserfromhtml/}

\bibitem{selectionBias}
Zadrozny, B.: Learning and evaluating classifiers under sample selection bias.
\newblock In: Proc. of the 21st Int. Conf. on Machine Learning, ICML '04, p.
  114. Association for Computing Machinery, New York, NY, USA (2004).
\newblock \doi{10.1145/1015330.1015425}.
\newblock \urlprefix\url{https://doi.org/10.1145/1015330.1015425}

\bibitem{10.1162/qss_a_00171}
Zagovora, O., Ulloa, R., Weller, K., Flöck, F.: {“I updated the
  $\langle$ref$\rangle$: The evolution of references in the English Wikipedia
  and the implications for altmetrics}.
\newblock Quantitative Science Studies \textbf{3}(1), 147--173 (2022).
\newblock \doi{10.1162/qss_a_00171}.
\newblock \urlprefix\url{https://doi.org/10.1162/qss\_a\_00171}

\bibitem{https://doi.org/10.1002/asi.24723}
Zheng, X., Chen, J., Yan, E., Ni, C.: Gender and country biases in {Wikipedia}
  citations to scholarly publications.
\newblock Journal of the Association for Information Science and Technology
  \textbf{74}(2), 219--233 (2023).
\newblock \doi{https://doi.org/10.1002/asi.24723}.
\newblock
  \urlprefix\url{https://asistdl.onlinelibrary.wiley.com/doi/abs/10.1002/asi.24723}

\end{thebibliography}
\end{document}